\documentclass[reprint, superscriptaddress, twocolumn, amsmath, amssymb, aps, prb]{revtex4-2}
\usepackage{graphicx}
\usepackage{dcolumn}
\usepackage{bm}
\usepackage{placeins}
\usepackage{appendix}
\usepackage{upgreek}

\usepackage{verbatim}
\usepackage[usenames,dvipsnames]{xcolor}
 \usepackage{amsmath}
 \usepackage{amsfonts}
 \usepackage{amssymb}
 \usepackage[colorlinks=true,citecolor=blue,linkcolor=red]{hyperref}

 \usepackage{bbold}     
 \usepackage[makeroom]{cancel}  
 \usepackage{multirow}    
 \usepackage[normalem]{ulem}        
 \usepackage{array}
 \usepackage{pdfpages}
\makeatletter
 \AtBeginDocument{\let\LS@rot\@undefined}
 \makeatother

\begin{document}


\title{Anisotropy of PbTe nanowires with and without a superconductor}


\author{Zonglin Li}
\email{equal contribution}
\affiliation{State Key Laboratory of Low Dimensional Quantum Physics, Department of Physics, Tsinghua University, Beijing 100084, China}

\author{Wenyu Song}
\email{equal contribution}
\affiliation{State Key Laboratory of Low Dimensional Quantum Physics, Department of Physics, Tsinghua University, Beijing 100084, China}

\author{Shan Zhang}
\email{equal contribution}
\affiliation{State Key Laboratory of Low Dimensional Quantum Physics, Department of Physics, Tsinghua University, Beijing 100084, China}

\author{Yuhao Wang}
\email{equal contribution}
\affiliation{State Key Laboratory of Low Dimensional Quantum Physics, Department of Physics, Tsinghua University, Beijing 100084, China}

\author{Zhaoyu Wang}
\email{equal contribution}
\affiliation{State Key Laboratory of Low Dimensional Quantum Physics, Department of Physics, Tsinghua University, Beijing 100084, China}

\author{Zehao Yu}
\affiliation{State Key Laboratory of Low Dimensional Quantum Physics, Department of Physics, Tsinghua University, Beijing 100084, China}

\author{Ruidong Li}
\affiliation{State Key Laboratory of Low Dimensional Quantum Physics, Department of Physics, Tsinghua University, Beijing 100084, China}

\author{Zeyu Yan}
\affiliation{State Key Laboratory of Low Dimensional Quantum Physics, Department of Physics, Tsinghua University, Beijing 100084, China}

\author{Jiaye Xu}
\affiliation{State Key Laboratory of Low Dimensional Quantum Physics, Department of Physics, Tsinghua University, Beijing 100084, China}

\author{Yichun Gao}
\affiliation{State Key Laboratory of Low Dimensional Quantum Physics, Department of Physics, Tsinghua University, Beijing 100084, China}

\author{Shuai Yang}
\affiliation{State Key Laboratory of Low Dimensional Quantum Physics, Department of Physics, Tsinghua University, Beijing 100084, China}

\author{Lining Yang}
\affiliation{State Key Laboratory of Low Dimensional Quantum Physics, Department of Physics, Tsinghua University, Beijing 100084, China}

\author{Xiao Feng}
\affiliation{State Key Laboratory of Low Dimensional Quantum Physics, Department of Physics, Tsinghua University, Beijing 100084, China}
\affiliation{Beijing Academy of Quantum Information Sciences, Beijing 100193, China}
\affiliation{Frontier Science Center for Quantum Information, Beijing 100084, China}
\affiliation{Hefei National Laboratory, Hefei 230088, China}

\author{Tiantian Wang}
\affiliation{Beijing Academy of Quantum Information Sciences, Beijing 100193, China}
\affiliation{Hefei National Laboratory, Hefei 230088, China}

\author{Yunyi Zang}
\affiliation{Beijing Academy of Quantum Information Sciences, Beijing 100193, China}
\affiliation{Hefei National Laboratory, Hefei 230088, China}

\author{Lin Li}
\affiliation{Beijing Academy of Quantum Information Sciences, Beijing 100193, China}
\affiliation{Hefei National Laboratory, Hefei 230088, China}

\author{Runan Shang}
\affiliation{Beijing Academy of Quantum Information Sciences, Beijing 100193, China}
\affiliation{Hefei National Laboratory, Hefei 230088, China}

\author{Qi-Kun Xue}
\affiliation{State Key Laboratory of Low Dimensional Quantum Physics, Department of Physics, Tsinghua University, Beijing 100084, China}
\affiliation{Beijing Academy of Quantum Information Sciences, Beijing 100193, China}
\affiliation{Frontier Science Center for Quantum Information, Beijing 100084, China}
\affiliation{Hefei National Laboratory, Hefei 230088, China}
\affiliation{Southern University of Science and Technology, Shenzhen 518055, China}

\author{Ke He}
\email{kehe@tsinghua.edu.cn}
\affiliation{State Key Laboratory of Low Dimensional Quantum Physics, Department of Physics, Tsinghua University, Beijing 100084, China}
\affiliation{Beijing Academy of Quantum Information Sciences, Beijing 100193, China}
\affiliation{Frontier Science Center for Quantum Information, Beijing 100084, China}
\affiliation{Hefei National Laboratory, Hefei 230088, China}

\author{Hao Zhang}
\email{hzquantum@mail.tsinghua.edu.cn}
\affiliation{State Key Laboratory of Low Dimensional Quantum Physics, Department of Physics, Tsinghua University, Beijing 100084, China}
\affiliation{Beijing Academy of Quantum Information Sciences, Beijing 100193, China}
\affiliation{Frontier Science Center for Quantum Information, Beijing 100084, China}

\begin{abstract}

We investigate the anisotropic behaviors in PbTe and PbTe-Pb hybrid nanowires. In previous studies on PbTe, wire-to-wire variations in anisotropy indicate poor device control, posing a serious challenge for applications. Here, we achieve reproducible anisotropy in PbTe nanowires through a substantial reduction of disorder. We then couple PbTe to a superconductor Pb, and observe a pronounced deviation in the anisotropy behavior compared to bare PbTe nanowires. This deviation is gate-tunable and attributed to spin-orbit interaction and orbital effect, controlled by charge transfer between Pb and PbTe. These results provide a guidance for the controlled engineering of exotic quantum states in this hybrid material platform.

\end{abstract}

\maketitle  

PbTe semiconductor nanowires have recently emerged as a promising material platform for the realization of Majorana zero modes \cite{Lutchyn2010, Oreg2010, CaoZhanPbTe, Jiangyuying, Erik_PbTe_SAG, PbTe_AB, Frolov_PbTe, Fabrizio_PbTe, Zitong, Wenyu_QPC, Yichun, Yuhao, Ruidong, Vlad_PbTe, Wenyu_Disorder, PbTe_In, Yuhao_degeneracy, Yichun_SQUID, Quantized_Andreev}. In contrast to well-studied systems such as InAs and InSb \cite{Mourik, Deng2016, Suominen_ZBP, Gul2018, Song2022, WangZhaoyu, Delft_Kitaev, MS_2023, NextSteps, Prada2020, Leo_perspective}, PbTe offers a significant advantage due to its large dielectric constant ($\sim$ 1350). This advantage can effectively mitigate disorder, the primary challenge in Majorana research \cite{Patrick_Lee_disorder_2012, Prada2012, Loss2013ZBP, GoodBadUgly, DasSarma_estimate, DasSarma2021Disorder, Tudor2021Disorder, Loss_Andreev_band}. A direct evidence is experimental observations of quantized conductance at zero magnetic field in long-channel devices (up to 1.7 $\upmu$m) \cite{Wenyu_Disorder}, as well as quantized Andreev conductance in PbTe-based hybrids \cite{Quantized_Andreev}, outperforming the best-optimized III-V nanowires.

While these advancements are encouraging, the anisotropic properties of PbTe introduce new complexities \cite{CaoZhanPbTe}. Take the electron Land\'{e}  $g$-factor for instance. The anisotropic $g$ tensor in PbTe complicates the interplay between Zeeman energy and spin-orbit interaction. A magnetic field aligned along one direction can induce additional Zeeman splitting in other directions.  Previous studies on PbTe \cite{Fabrizio_PbTe} have shown that the $g$-factor anisotropy is device dependent, exhibiting no consistent relationship with the crystal orientation. This ``randomness'' poses a significant hurdle for applications aimed at engineering Majorana states or other exotic quantum phenomena. Recently, we have reduced disorder in PbTe nanowires \cite{Wenyu_Disorder}, allowing us to achieve better device control. Here, we report quantum dots defined in PbTe that exhibit reproducible $g$-factor anisotropy, consistent across devices. We also explore the effects of introducing a superconductor into the PbTe-Pb hybrids. The anisotropy deviates from that of pure PbTe. The deviation is gate tunable and can be attributed to charge transfer, spin-orbit interaction and orbital effect. These findings provide a deeper understanding of this hybrid material system and offer valuable insights for controlled design of quantum devices.

\begin{figure}[htb]
\includegraphics[width=\columnwidth]{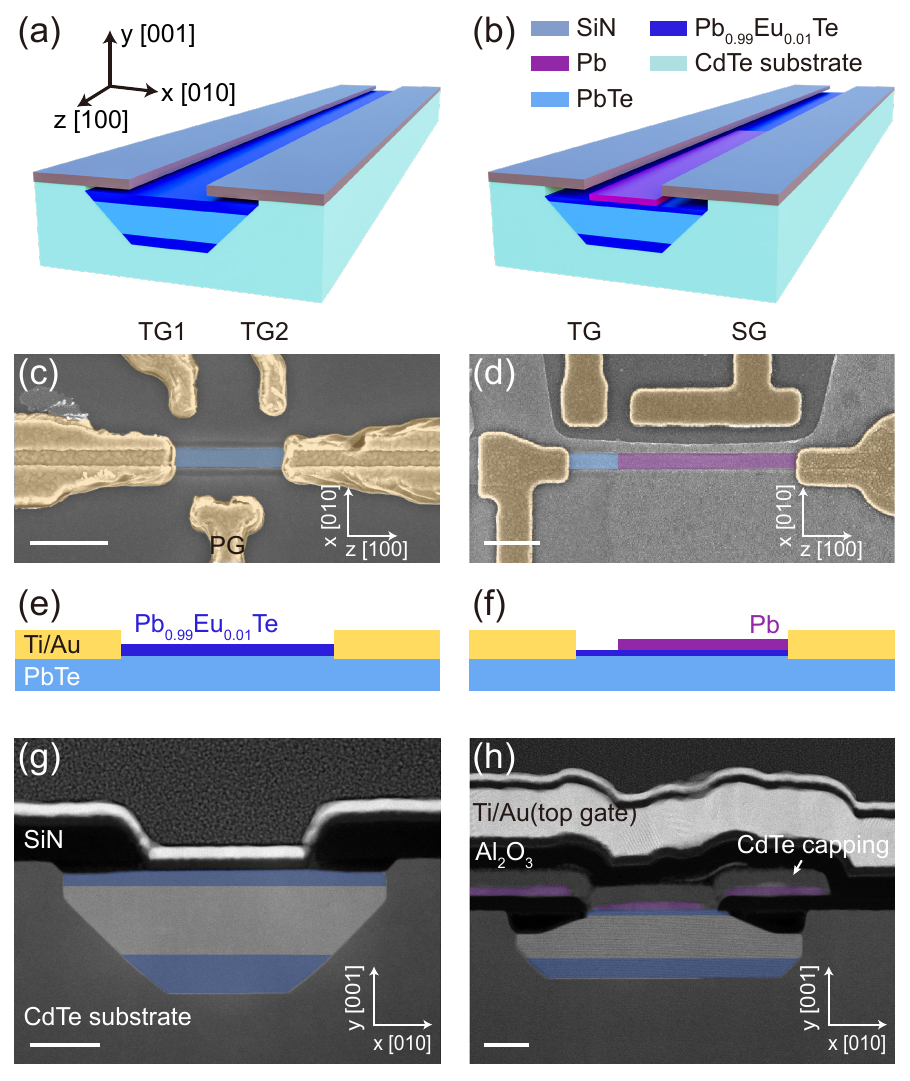}
\centering
\caption{Device introduction. (a-b) Brief schematics of PbTe and PbTe-Pb nanowires, respectively. The coordinate system is labeled. (c-d) False-colored SEMs of a typical PbTe and PbTe-Pb device. Scale bars, 500 nm. (e-f) Brief schematics along the longitudinal directions of the devices. (g-h) Cross-sectional STEMs. Scale bars, 50 nm. The PbEuTe regions are false-colored blue. The Pb film is in violet. }
\label{fig1}
\end{figure}

Figures 1(a-b) provide brief schematics of PbTe and PbTe-Pb wires. The coordinate system for all devices is defined as follows: the $z$-axis aligns with the longitudinal direction of the wire, the $x$-axis is in-plane and perpendicular to the wire, and the $y$-axis is out-of-plane \cite{axis}. These wires were selectively grown on a CdTe(100) substrate, with the longitudinal direction aligned along the [100] crystal axis \cite{Wenyu_Disorder}. For hybrid wires, a thin Pb$_{0.99}$Eu$_{0.01}$Te interlayer, with a thickness ranging from 4 to 9 nm, was grown between PbTe and Pb to tune induced superconductivity. Figure 1(c) shows a scanning electron micrograph (SEM) of a PbTe device, where three side gates (TG1, TG2, and PG) are used to define a quantum dot. Figure 1(d) is an SEM of a PbTe-Pb device, in which two gates, TG and SG, were fabricated to tune the tunnel barrier region and the superconducting proximitized region. Some PbTe-Pb nanowires are equipped with top gates, while others use side gates for control. Figures 1(e-f) show schematics of the devices along the longitudinal direction. In Figs. 1(b) and 1(f), the CdTe capping layer has been omitted for clarity. Device fabrication follows that in our previous works \cite{Wenyu_Disorder}.

We conducted scanning transmission electron microscopy (STEM) on each device after its measurement. Figures 1(g-h) present two representative examples. Embedding PbTe into the CdTe substrate results in an environment with matching lattice constants. Further capping the wires with CdTe or Pb$_{0.99}$Eu$_{0.01}$Te serves to eliminate surface oxides from the core device region. These improvements are the crucial to achieve controllable device performance. Figure 1(h) depicts a top-gate device. The top gate, composed of Ti/Au, is separated from the device by a dielectric layer of Al$_2$O$_3$, which is grown via atomic layer deposition. In this study, a total of 9 devices were analyzed: 3 PbTe and 6 PbTe-Pb nanowires. For a comprehensive overview of their SEMs and STEMs, we refer to Fig. S1 in the Supplemental Material (SM). All measurements were conducted in a dilution refrigerator at a base temperature of less than 50 mK, with a standard two-terminal measurement configuration \cite{Zitong}.

\begin{figure}[htb]
\includegraphics[width=\columnwidth]{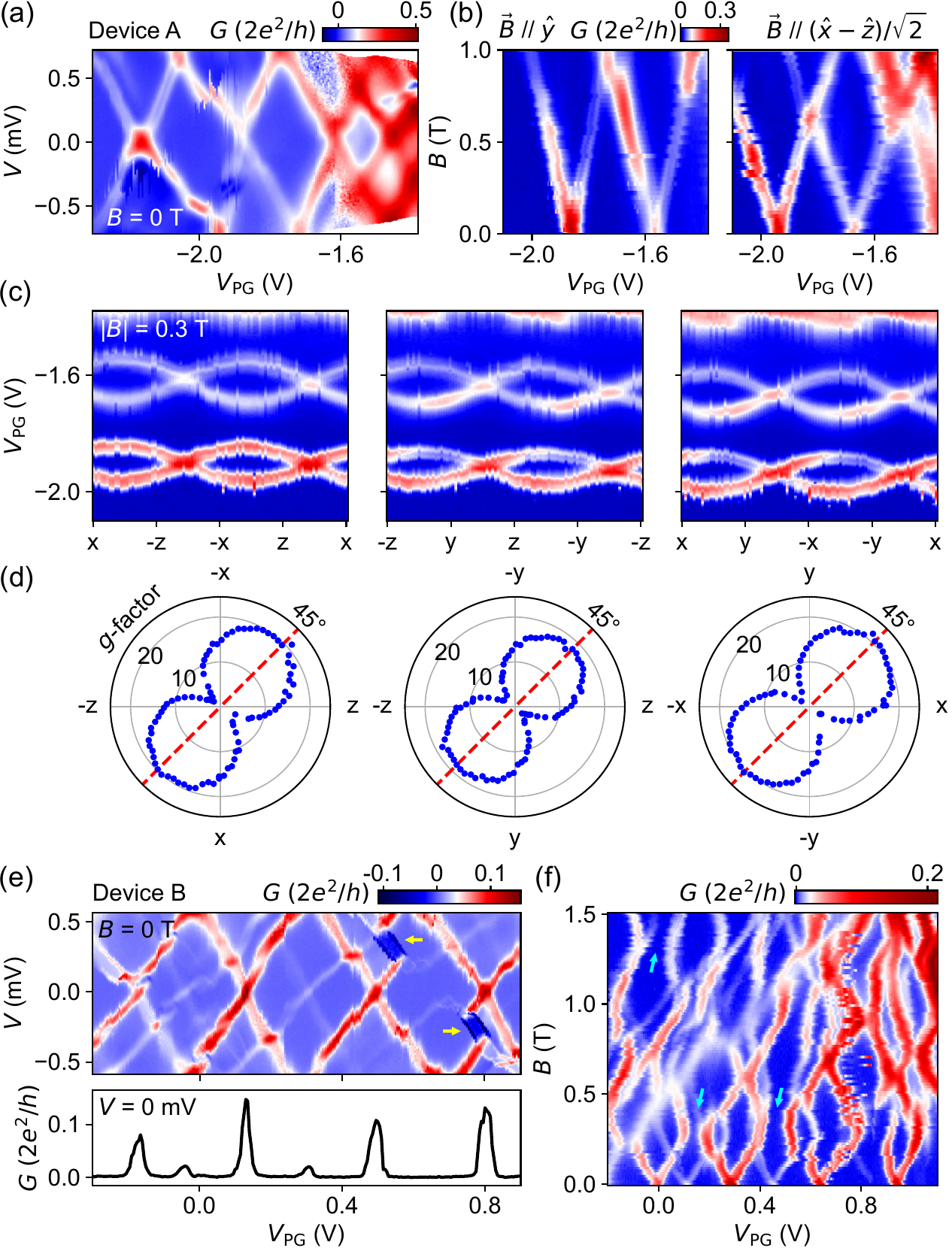}
\centering
\caption{Anisotropy in PbTe nanowires. (a) Charge stability diagram of device A at 0 T. (b) Field dependence of dot levels in device A. $B$ directions are labeled. $V$ = 0 mV. (c) Angle dependence of the level splitting by rotating $B$ in three planes. $|B|$ = 0.3 T. $V_{\text{TG1}}$ = -2 V, $V_{\text{TG2}}$ = -1.92 V for (b-c). (d) Polar plots of $g$-factors extracted from (c). (e) Charge stability diagram of device B at 0 T. Lower panel, zero-bias line cut. (f) $B$ dependence of dot levels in device B. $V$ = 0 mV. $B$ direction is $(-\hat{x}+\hat{z})/\sqrt{2}$.  }
\label{fig2}
\end{figure}

We first examine the PbTe devices. Figure 2(a) shows the charge stability diagram of device A, measured at zero magnetic field. The tunnel gate voltages, $V_{\text{TG1}}$ and $V_{\text{TG2}}$, were fixed at -2 V and -1 V, respectively, while the plunger gate voltage $V_{\text{PG}}$ was varied. $V$ is the bias voltage drop across the device. Differential conductance, $G \equiv dI/dV$, is plotted. Diamond-like structures are revealed, reminiscent of Coulomb blockades. When a magnetic field $B$ is applied (Fig. 2(b)), the conductance peaks split, suggesting negligible charging energy in this system. This observation is consistent with previous reports \cite{Frolov_PbTe, Fabrizio_PbTe}. The origin of small or negligible charging energy is attributed to the large dielectric constant of PbTe, which effectively screens electrostatic interactions. Consequently, the diamonds in Fig. 2(a) correspond to quantized levels formed in the PbTe quantum dot, rather than Coulomb charging energy. The level spacing, estimated from the diamond size, is $\sim$ 0.6 meV, in good agreement with values reported in Ref.\cite{Fabrizio_PbTe}.

Figure 2(b) shows the Zeeman splitting of two adjacent levels. $B$ was aligned in two directions: $\hat{y}$ and $(\hat{x}-\hat{z})/\sqrt{2}$. In the $\hat{y}$ direction, the two neighboring levels with opposite spins cross near 1 T, while for the other direction, the crossing occurs at $\sim$ 0.6 T. This difference indicates $g$-factor anisotropy. Figure S2 in SM shows additional $B$ scans along other directions. To map the anisotropy in all directions, we fix the magnitude of $B$ at 0.3 T and rotate its direction. Figure 2(c) illustrates the rotation in three planes: $xz$, $yz$, and $xy$. The left panel corresponds to the $xz$ plane, which is parallel to the device substrate. The peak splitting is minimized (nearly non-split) for $B$ aligned along $\pm(\hat{x}+\hat{z})/\sqrt{2}$, i.e. 45$^{\circ}$ between the $x$ and $z$ axes. This direction is associated with the minimum $g$-factor (close to zero). In contrast, the $g$-factor and peak splitting are maximized when $B$ is aligned along $\pm(-\hat{x}+\hat{z})/\sqrt{2}$, i.e. 45$^{\circ}$ to the $z$ and $-x$ axes. A similar anisotropy is observed in the other two planes.

For a quantitative analysis, we first determine the lever arm, $\alpha \sim$ 2.7 meV/V, which relates the energy scale to the gate voltage $V_{\text{PG}}$, based on the diamond size in Fig. 2(a). Using this lever arm, we convert the peak splitting in Fig. 2(c) from $V_{\text{PG}}$ to energy, and then divide it by $\mu_{\text{B}}B$ to extract the $g$-factor \cite{error}. The results are shown in Fig. 2(d) using polar plots. The $g$-factor ranges from 0 to 21. Notably, the directions of the maximum and minimum $g$-factor do not align with the wire axis (or the crystal axis [100]) but are instead $\sim$ 45$^{\circ}$ offset. For a guidance, we mark 45$^{\circ}$ direction to the $-x$ and $z$ axes with a red dashed line. This 45$^{\circ}$ offset differs from previous reports on nanowires with the same crystal orientation \cite{Frolov_PbTe}. Our anisotropic pattern is reproducible across multiple devices, as shown in Figs. S3 and S4. While the magnitude of $g$-factor can vary between devices due to differences in wire thickness, the anisotropic direction remains consistent and reproducible. This consistency addresses the issue of device-to-device variation reported in earlier studies \cite{Fabrizio_PbTe}. Our improved control on device performance is crucial for designing more advanced devices incorporating superconductors.

\begin{figure}[htb]
\includegraphics[width=\columnwidth]{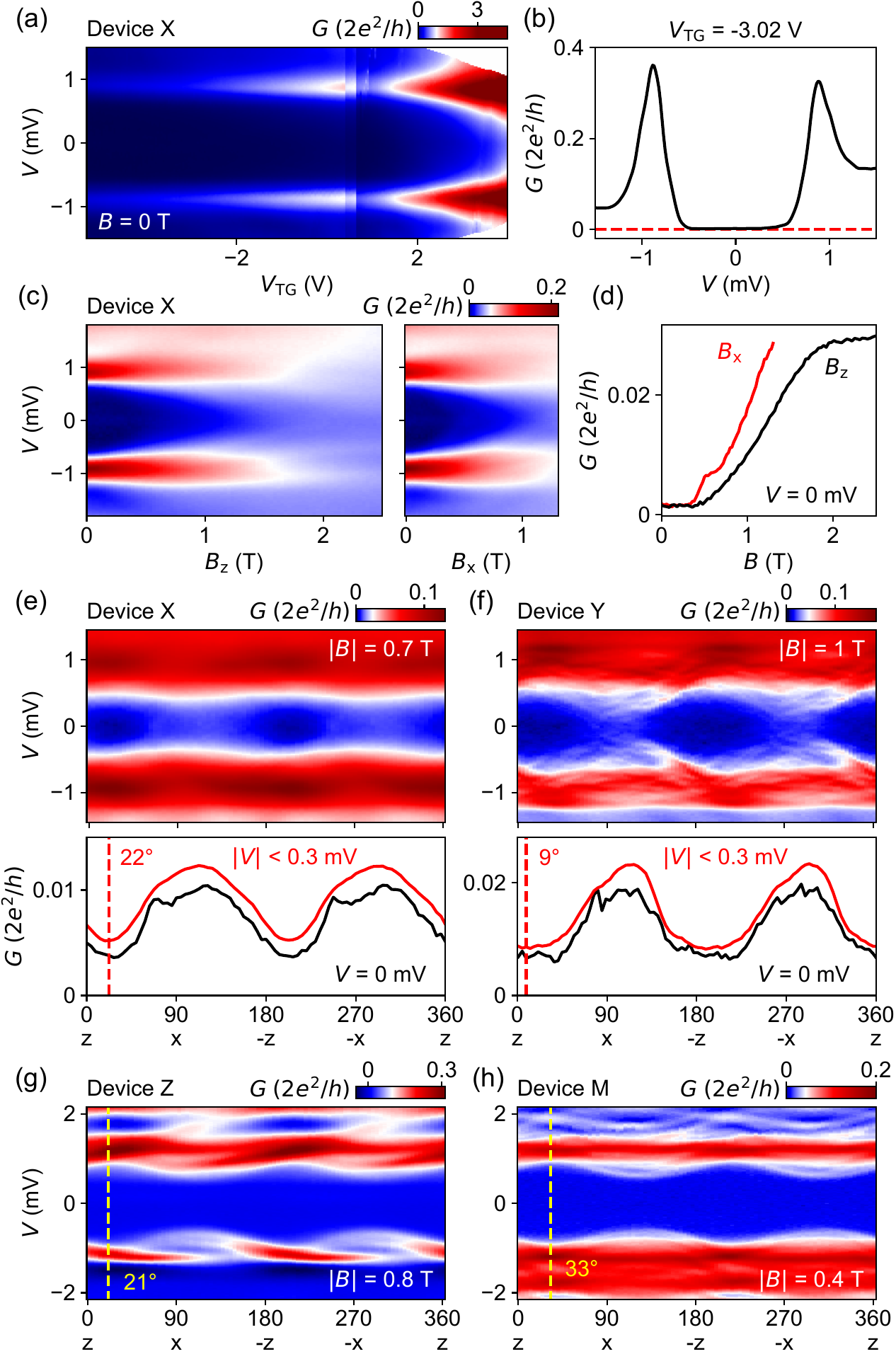}
\centering
\caption{Anisotropy in PbTe-Pb nanowires. (a) Gap spectroscopy of device X at 0 T. $V_{\text{SG}} =$ -2.5 V for (a-e). (b) A line cut from (a) at $V_{\text{TG}} = -3.02$ V.  (c) $B$ scans of the gap along $z$ and $x$ axes. $V_{\text{TG}}$ = -4.8 V and -4.68 V, respectively. (d) Zero-bias line cuts from (c). (e) Angle dependence of the gap in device X. $|B|$ = 0.7 T. $V_{\text{TG}}$ = -4.3 V. Lower panel, zero-bias line cut (black) and averaged $G$ for $|V| < 0.3$ mV (red). (f-h) Angle dependence of gap anisotropy in three additional devices. From (f) to (h): $V_{\text{TG}}$ = -0.98 V,  1.06 V, -2.07 V, and $V_{\text{SG}}$  = 0 V, 2 V, -2.5 V.  }
\label{fig3}
\end{figure}

Figure 2(e) displays the charge stability diagram of a second device, with the lower panel showing the zero-bias line cut. An ``even-odd'' pattern is observed in the peak heights. Additionally, negative differential conductance appears at finite biases in certain gate voltage regions, as marked by the yellow arrows. The origin of these phenomena remains unclear. Nevertheless, all the peaks split upon increasing $B$, as shown in Fig. 2(f). The zero-field peaks, therefore, correspond to spin-degenerate levels formed within the dot. As increasing $B$, peaks from different levels with opposite spins approach each other and eventually exhibit avoided crossings, indicated by the cyan arrows. These anti-crossings, also observed in device A (Fig. S2), arise from the spin-orbit interaction, where states with opposite spins are no longer orthogonal. For additional data and analysis on device B, as well as a third device (device C), see Figs. S3 and S4.

Building upon the systematic understanding of anisotropy in PbTe, we extend our study to PbTe-Pb hybrids to investigate the impact of superconductivity on anisotropy.  Figure 3(a) shows the tunneling spectroscopy of a PbTe-Pb nanowire (device X). A clean and hard superconducting gap is observed, see Fig. 3(b) for a line cut. The gap size, $\Delta \sim$ 0.9 meV,  is slightly smaller than the bulk gap of Pb ($\sim$ 1 meV). This reduction is achieved by the presence of a 4-nm-thick Pb$_{0.99}$Eu$_{0.01}$Te interlayer between PbTe and Pb, which modulates the PbTe/Pb coupling strength. The jumps near $V_{\text{TG}} = 0.5$ V are due to charge instabilities. Figure 3(c) shows $B$ scans along the $z$ and $x$ axes. The critical field of the gap is larger for $B_z$ than for $B_x$. This anisotropy is better revealed in the zero-bias line cuts shown in Fig. 3(d).

We next fix the magnitude of $B$ to 0.7 T, and rotate its direction within the $xz$ plane. Figure 3(e) shows this rotation.  The lower panel displays the zero-bias line cut (black) and the conductance averaged near zero bias ($|V| <$ 0.3 mV, red). Both curves consistently reveal anisotropy in the gap softness. The minimum conductance corresponds to an angle of 22$^{\circ}$ to the $z$ axis. This direction differs significantly from the 45$^{\circ}$ observed in bare PbTe nanowires. The 23$^{\circ}$ deviation is attributed to the presence of the superconductor. Note that the $xz$ plane is parallel to the substrate, where the orbital effects of the Pb film are negligible for $B$ aligned in this plane. If the Zeeman effect alone were responsible for the gap closure, the gap anisotropy in the $xz$ plane would follow the $g$-factor anisotropy observed in PbTe wires. However, the 23$^{\circ}$ deviation suggests that the presence of Pb imposes other mechanisms affecting the gap closure such as spin-orbit interaction.

The interplay between spin-orbit interaction and Zeeman energy has been studied in InSb-NbTiN hybrid nanowires \cite{Jouri2019}. When $B$ is perpendicular to the spin-orbit direction, the critical field for gap closure is larger, as the spin-orbit interaction protects the gap from being closed \cite{Lutchyn2010, Oreg2010}. In the case of InSb-NbTiN, the direction of minimal zero-bias conductance at finite $B$ aligns with the nanowire axis (0$^{\circ}$ to the $z$ axis). However, in PbTe-Pb systems, this relationship deviates due to the complexity of substantial $g$-factor anisotropy. The observed 22$^{\circ}$ can be interpreted as the result of competing effects between spin-orbit interaction and the anisotropy of the $g$-factor (Zeeman effect). The former ``pushes'' this direction toward 0$^{\circ}$, while the latter ``prefers'' 45$^{\circ}$ to the $z$ axis. Note that the strength of spin-orbit interaction is affected by many factors. For instance, the coupling and charge transfer between PbTe and Pb can affect the electric field, which in turn affect the spin-orbit interaction. This can be tuned by the thickness of Pb$_{0.99}$Eu$_{0.01}$Te interlayer between PbTe and Pb.  Additionally, the PbTe thickness can be another factor. Figures 3(f-h) show three additional devices with different thicknesses in PbTe and Pb$_{0.99}$Eu$_{0.01}$Te interlayer (see Fig. S1 for their STEMs). The directions corresponding to the minimal zero-bias conductance are 9$^{\circ}$, 21$^{\circ}$, and 33$^{\circ}$, respectively. These variations reflect the ongoing competition between spin-orbit interaction and the Zeeman effect across devices.

\begin{figure}[htb]
\includegraphics[width=\columnwidth]{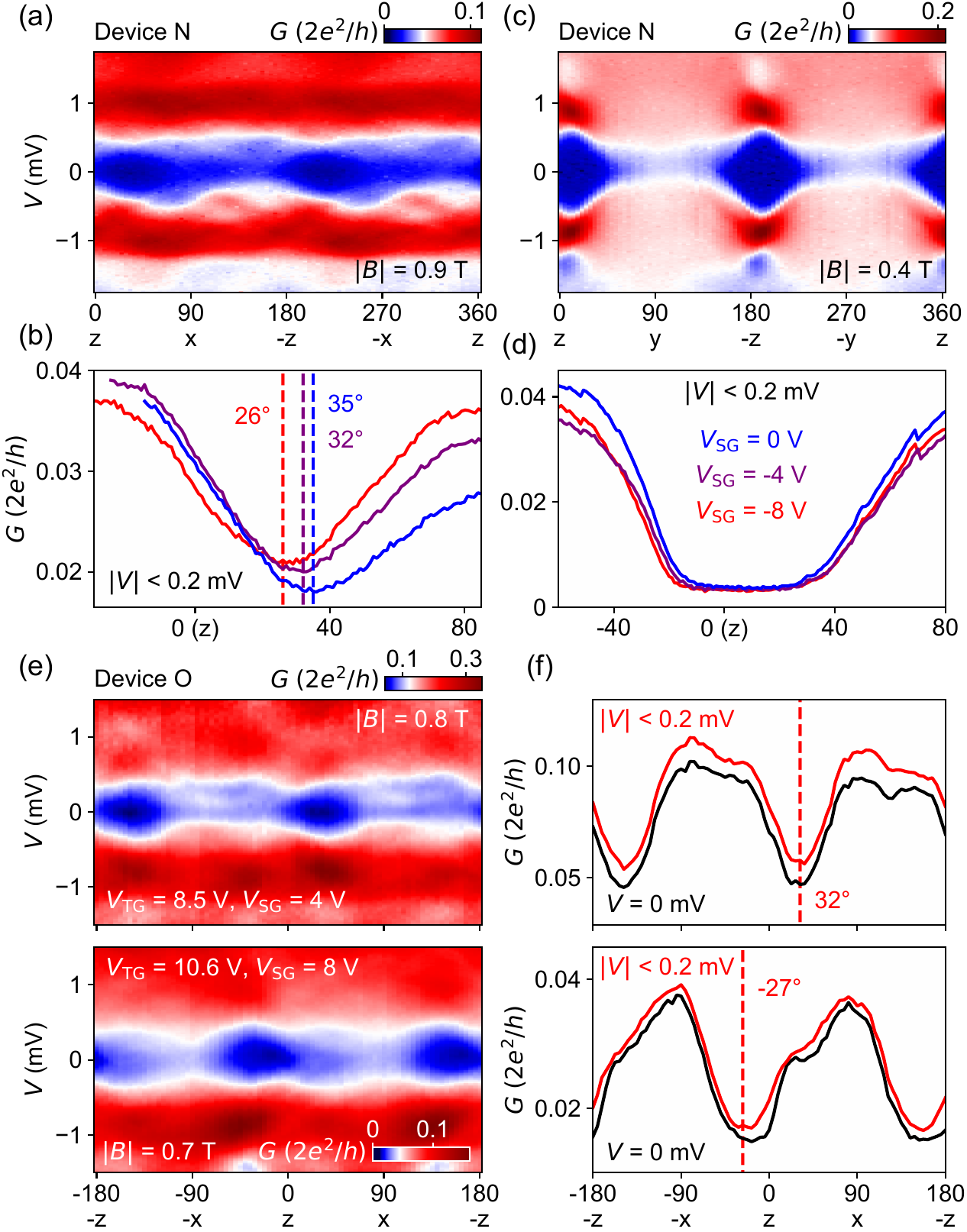}
\centering
\caption{Gate-tunable anisotropy in PbTe-Pb. (a) Gap anisotropy in device N at $V_{\text{SG}}$ = 0 V.  (b) Averaged conductance for $|V| <$ 0.2 mV. $V_{\text{SG}}$ for the three curves are 0 V (blue), -4 V (violet), and -8 V (red), respectively. $V_{\text{TG}}$ = -2.2 V, -1.2 V, and -1.32 V.  (c-d) Angle dependence of device N in the $yz$ plane. (e) Gap anisotropy in device O at $V_{\text{SG}}$ = 4 V (upper) and 8 V (lower), respectively. (f) Zero-bias line cuts (black) and averaged conductance near zero bias ($|V| <$ 0.2 mV) extracted from (e). }
\label{fig2}
\end{figure}

For out-of-plane $B$ (the $y$ axis), orbital effect becomes dominant. The area of Pb film perpendicular to $B$ is non-negligible, leading to a much smaller critical field compared to the $xz$-plane case. Consequently, $B$ rotations in $xy$ and $yz$ planes reveal that the direction of minimum zero-bias conductance is close to the $y$ axis, see Figs. S5 and S6 for these rotations on devices X, Y, Z, and M. 

Figure 4 presents the gate dependence of the anisotropy in PbTe-Pb hybrids. In Fig. 4(a), we rotate $B$ in the $xz$ plane for device N by fixing $V_{\text{SG}}$ to 0 V. The minimum zero-bias conductance corresponds to an angle of 35$^{\circ}$ to the $z$ axis. To better visualize this angle, we performed a fine measurement near this direction. The conductance averaged near zero bias ($|V| <$ 0.2 mV) is plotted as the blue curve in Fig. 4(b). The blue dashed line highlights the position of 35$^{\circ}$. The asymmetry of the curve with respect to 35$^{\circ}$ is likely caused by orbital effect in PbTe, a phenomenon also observed in InSb-NbTiN systems \cite{Jouri2019}. The violet and red curves in Fig. 4(b) correspond to $V_{\text{SG}}$ = -4 V and -8 V, respectively.  The dashed lines mark the corresponding angles of minimum conductance, which shift to 32$^{\circ}$ and 26$^{\circ}$. This 9$^{\circ}$ shift reflects the $V_{\text{SG}}$ tuning of spin-orbit interaction. It also addresses a long-standing concern regarding whether the large dielectric constant of PbTe screens and limits the tunability of $V_{\text{SG}}$ in the proximitized region.   

Figure 4(c) shows $B$ rotation in the $yz$ plane, revealing the direction of minimum conductance close to $z$ axis ($\sim$ 10$^{\circ}$). This direction is barely tuned by $V_{\text{SG}}$, as shown in Fig. 4(d), indicating that the orbital effect dominates the gap closure for non-zero out-of-plane $B$ components.  Figures 4(e-f) show another device revealing larger gate tunability of anisotropy in the $xz$ plane: from 32$^{\circ}$ to -27$^{\circ}$ for $V_{\text{SG}}$. For rotations in other planes, see Fig. S7.

To conclude, we have systematically studied the anisotropic behaviors in PbTe and PbTe-Pb nanowires. For bare PbTe nanowires, we identified $\pm$45$^{\circ}$ to the wire axis ([100]) as the direction of minimum and maximum $g$-factors. Coupling PbTe to Pb introduces deviations from these directions, highlighting the significant role of spin-orbit interaction due to the presence of Pb. The spin-orbit interaction can be further tuned by gate voltages and the Pb$_{0.99}$Eu$_{0.01}$Te interlayer thickness. These findings offer valuable guidance for the design of advanced quantum devices.

\section{Acknowledgment}

This work is supported by National Natural Science Foundation of China (92065206) and the Innovation Program for Quantum Science and Technology (2021ZD0302400). Raw data and processing codes within this paper are available at https://doi.org/10.5281/zenodo.14614293
 
\bibliography{mybibfile}

\begin{thebibliography}{41}
\bibitem [{axi()}]{axis}%
  \BibitemOpen
  \href@noop {} {\bibinfo {title} {There is an uncertainty on the orientation
  of $z$ and $-z$, which cannot be determined from {SEM} and {STEM}. the same
  also applies to $x$ and $-x$, as well as $y$ and $-y$.}}\BibitemShut {Stop}%
\end{thebibliography}

\newpage

\onecolumngrid

\newpage


\section*{Supplementary material}

# Supplemental Material for "Anisotropy of PbTe nanowires with and without a superconductor"

Zonglin Li,[1, *] Wenyu Song,[1, *] Shan Zhang,[1, *] Yuhao Wang,[1, *] Zhaoyu Wang,[1, *] Zehao Yu,[1] Ruidong Li,[1] Zeyu Yan,[1] Jiaye Xu,[1] Yichun Gao,[1] Shuai Yang,[1] Lining Yang,[1] Xiao Feng,[1, 2, 3, 4] Tiantian Wang,[2, 4] Yunyi Zang,[2, 4] Lin Li,[2] Runan Shang,[2, 4] Qi-Kun Xue,[1, 2, 3, 4, 5] Ke He,[1, 2, 3, 4, †] and Hao Zhang[1, 2, 3, ‡]

[1]*State Key Laboratory of Low Dimensional Quantum Physics,
Department of Physics, Tsinghua University, Beijing 100084, China*
[2]*Beijing Academy of Quantum Information Sciences, Beijing 100193, China*
[3]*Frontier Science Center for Quantum Information, Beijing 100084, China*
[4]*Hefei National Laboratory, Hefei 230088, China*
[5]*Southern University of Science and Technology, Shenzhen 518055, China*

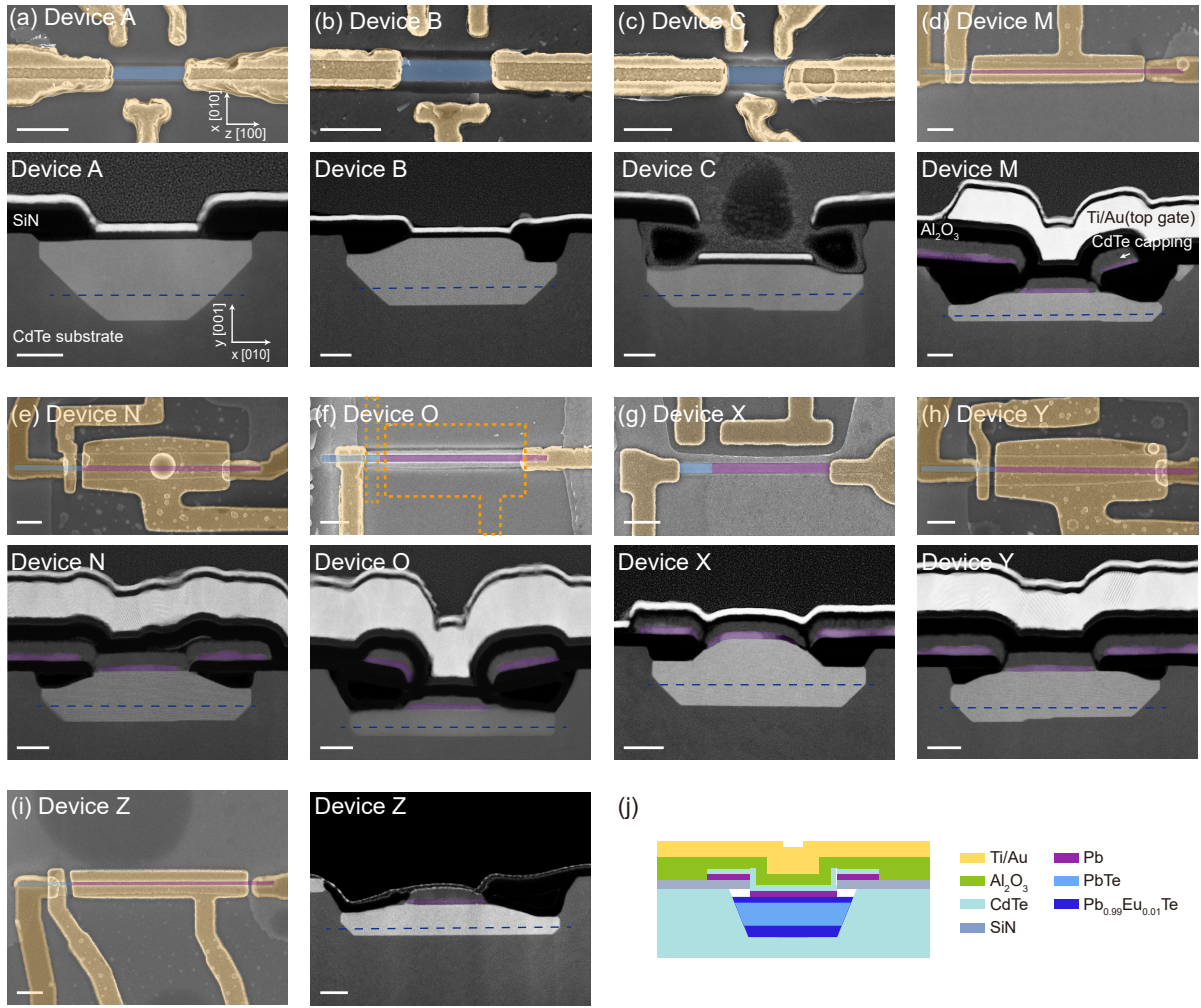

FIG. S1. SEMs (upper) and STEMs (lower) of the 9 devices. Scale bars, 500 nm for SEMs and 50 nm for STEMs. (a-c) Quantum dot devices. Blue dashed lines correspond to the interface between $Pb_{0.99}Eu_{0.01}Te$ buffer and PbTe. (d-i) PbTe-Pb devices. Device X has side gates while the rest devices have top gates. For device O, the SEM was taken before gate deposition. The orange dashed lines mark the edges of TG and SG. The Pb films are false-colored violet while PbTe is in blue. (j) Cross-sectional schematic of a PbTe-Pb device with top gates.

---

* equal contribution
† kehe@tsinghua.edu.cn
‡ hzquantum@mail.tsinghua.edu.cn

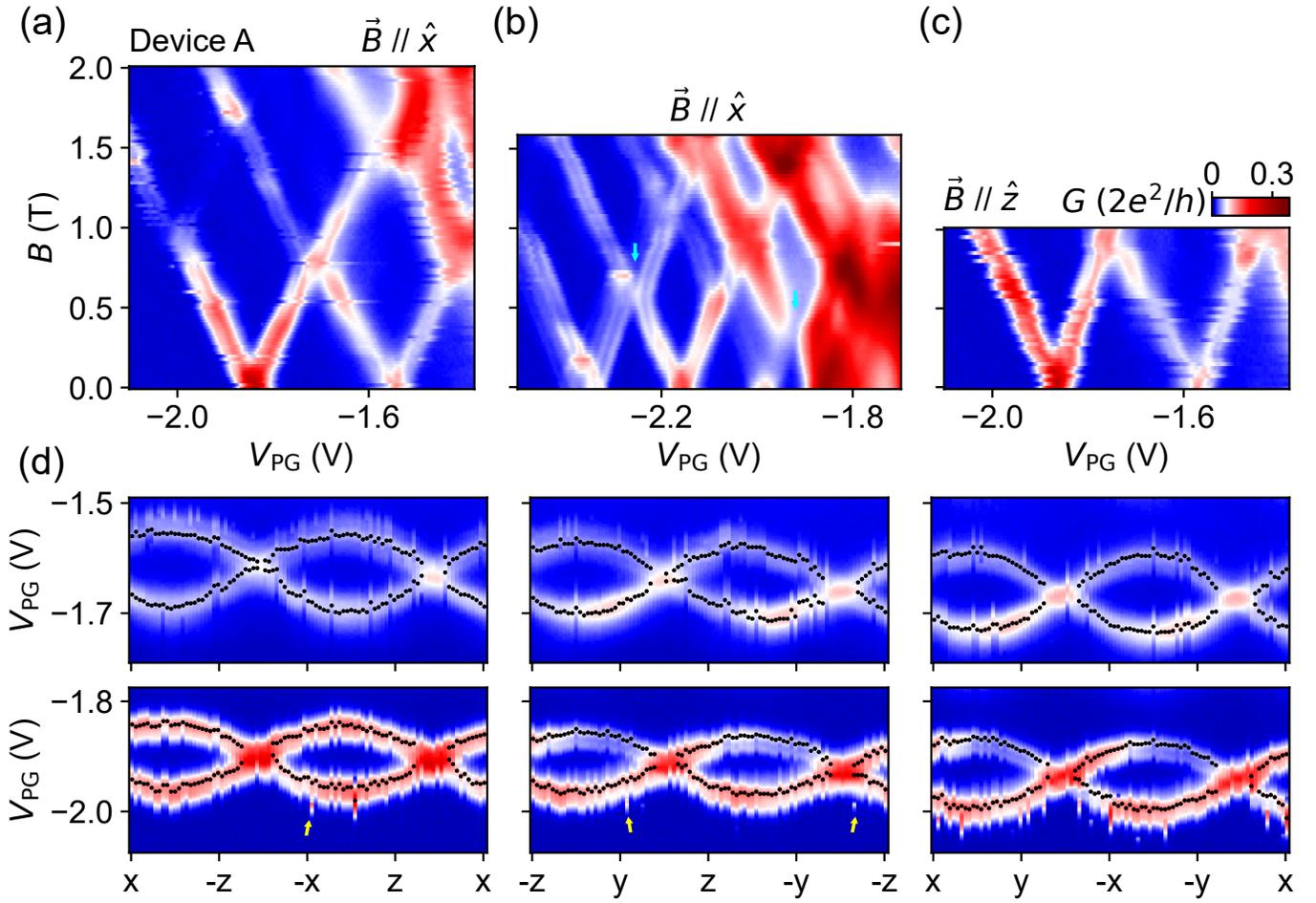

FIG. S2. Additional data of device A. (a-b) $B$ scans of dot levels along $x$ axis. $V_{TG1} = -2$ V, $V_{TG2} = -1.92$ V (a) and $-0.81$ V (b). The cyan arrows indicate level anti-crossings. (c) $B$ scan along the $z$ axis. (d) Extraction of the peak splittings in Fig. 2(c). At each fixed direction, we first used a "peak finding" method to locate the positions of the two split peaks. The peaks due to noise or jumps are discarded (see the yellow arrows for example). The black dots denotes the peak positions based on which we estimate the $g$ factor.

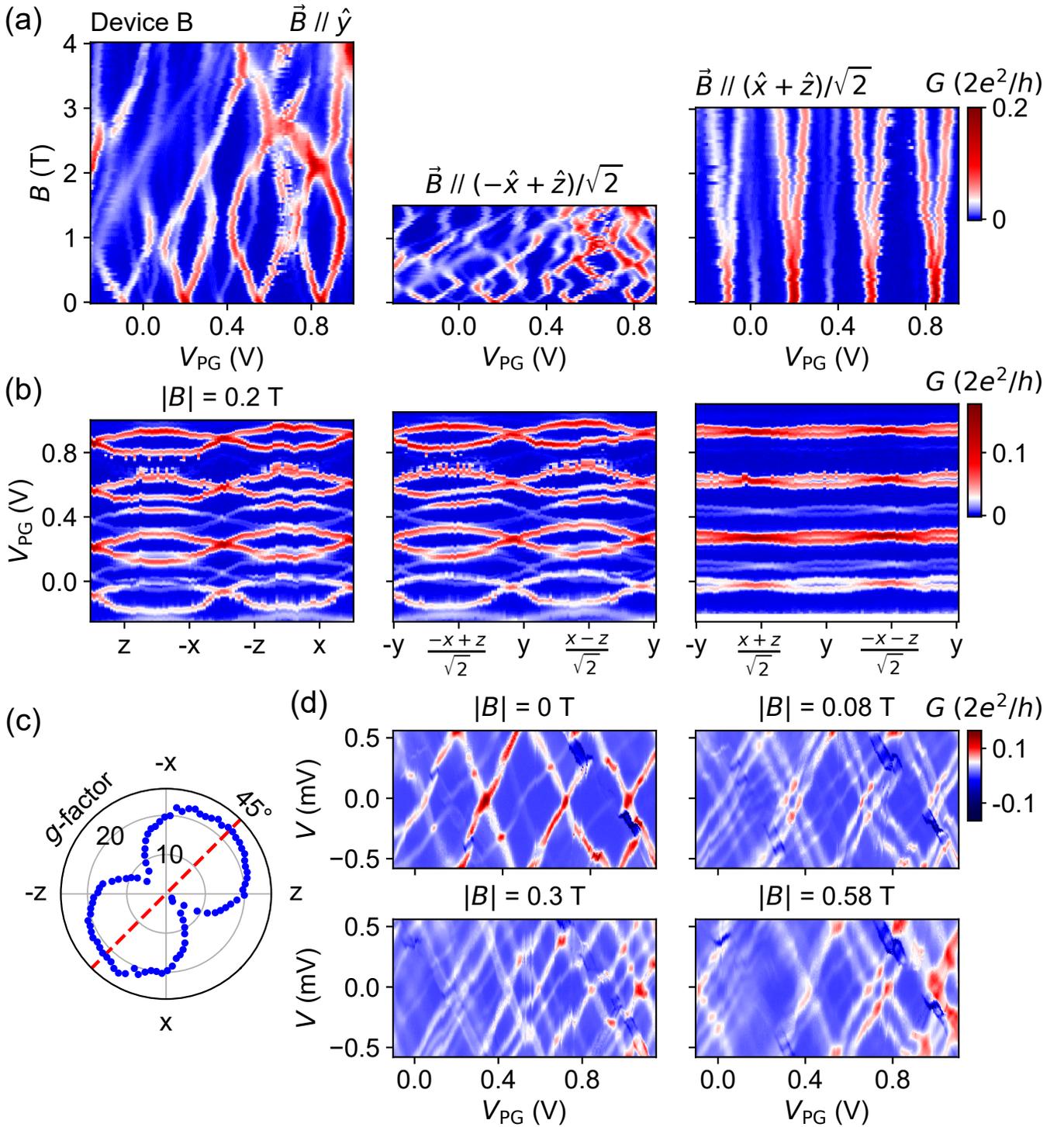

FIG. S3. Additional data of device B. (a) $B$ scans along the three directions: $\hat{y}$, $(-\hat{x}+\hat{z})/\sqrt{2}$, and $(\hat{x}+\hat{z})/\sqrt{2}$. Zeeman splittings with $g$-factor anisotropy and anti-crossings of dot levels can be observed. $V_{TG1} = -2.55$ V, $V_{TG2} = -2.55$ V. (b) Rotations in the three planes with a fixed magnitude of $|B| = 0.2$ T. The left panel corresponds to the $xz$ plane. The plane of the middle panel is defined by $\hat{y}$ and $(-\hat{x}+\hat{z})/\sqrt{2}$, i.e. different from the case of device A. The same also holds for the right panel. (c) Polar plots of $g$ factors in the $xz$ plane, extracted from (b). The red dashed line marks 45°. The $g$ factor may be underestimated near 45° due to level repulsion. (d) Charge stability diagram measured at different fields. The field direction is $(-\hat{x}+\hat{z})/\sqrt{2}$.

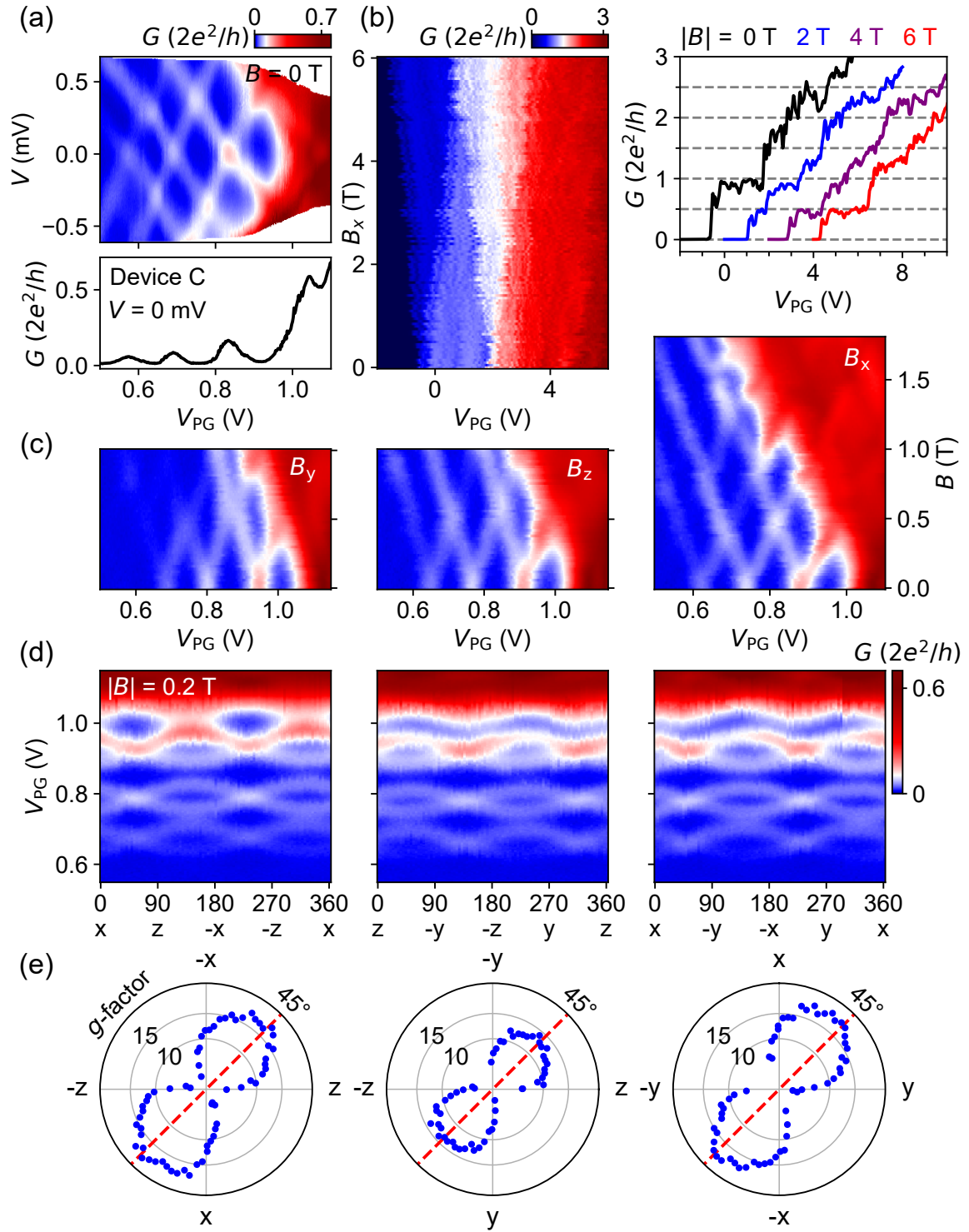

FIG. S4. Data of device C. (a) Charge stability diagram at zero field. $V_{\mathrm{TG1}} = -0.5$ V, $V_{\mathrm{TG2}} = -1.8$ V for (a-d). Lower panel, zero-bias line cut. (b) Quantized conductance in the open regime. The right panel shows line cuts at different fields with horizontal offsets for clarity. (c) $B$ scans of dot states along the three axes. (d) $B$ rotations in the three planes. $|B| = 0.2$ T. (e) $g$-factors extracted from (d). The anisotropic behavior is similar to that of device A.

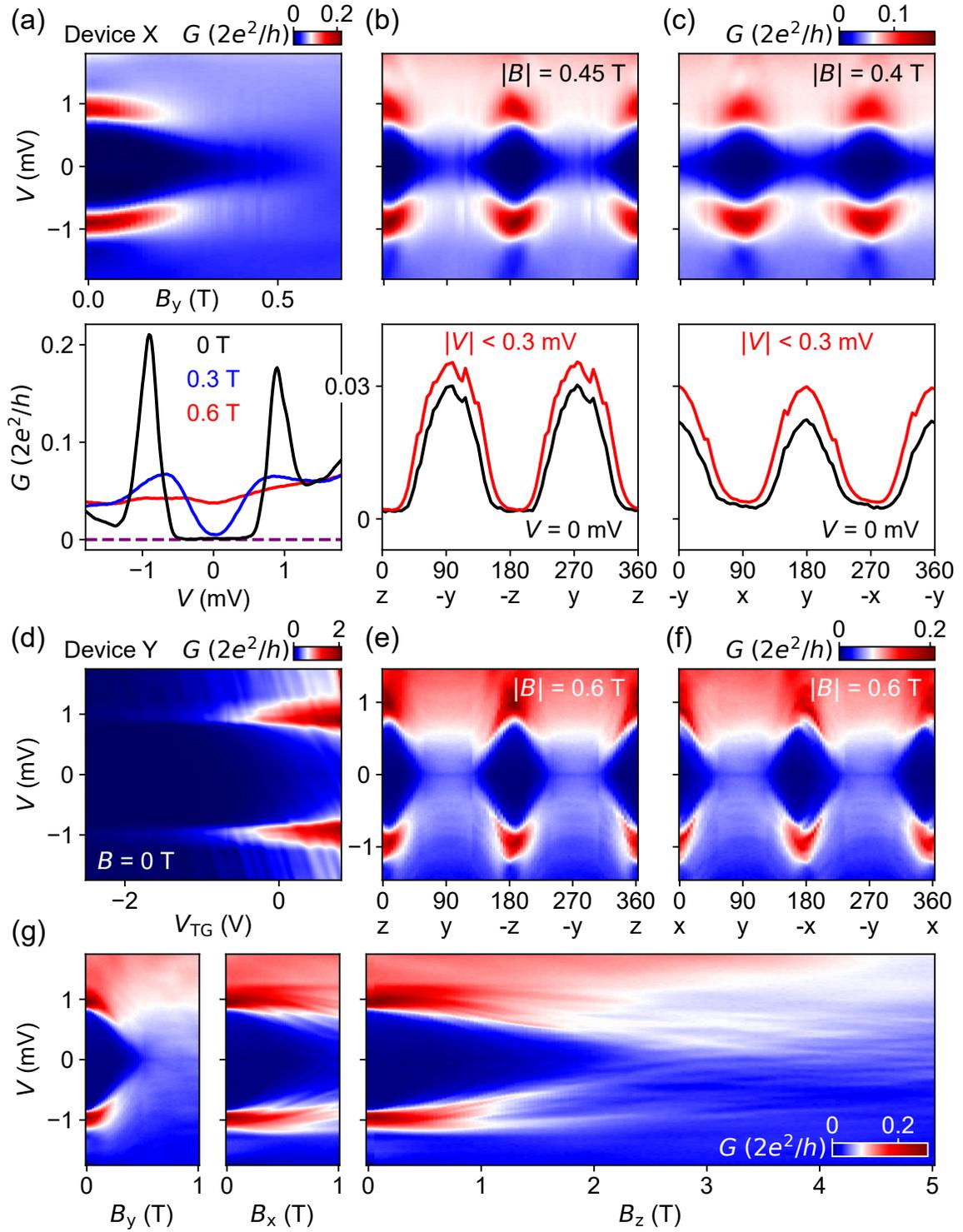

FIG. S5. Additional data of devices X (a-c) and Y (d-g). (a) $B_y$ scan of the gap of device X. $V_{TG}$ = -4.73 V, $V_{SG}$ = -2.5 V. Lower panel, three line cuts at 0, 0.3 T and 0.6 T. (b-c) $B$ rotations in the $yz$ plane and $xy$ plane. $V_{TG}$ = -4.3 V, $V_{SG}$ = -2.5 V. $|B|$ = 0.45 T (b) and 0.4 T (c). Lower panels, zero-bias line cuts (black) and average conductance for $|V| < 0.3$ mV (red). (d) Gap spectroscopy of device Y at zero field. $V_{SG}$ = 0 V. (e-f) $B$ rotations in the $yz$ and $xy$ planes. $|B|$ = 0.6 T. $V_{TG}$ = -0.98 V, $V_{SG}$ = 0 V. (g) $B$ scans of the gap for $B$ aligning with $y$, $x$, and $z$ axes.

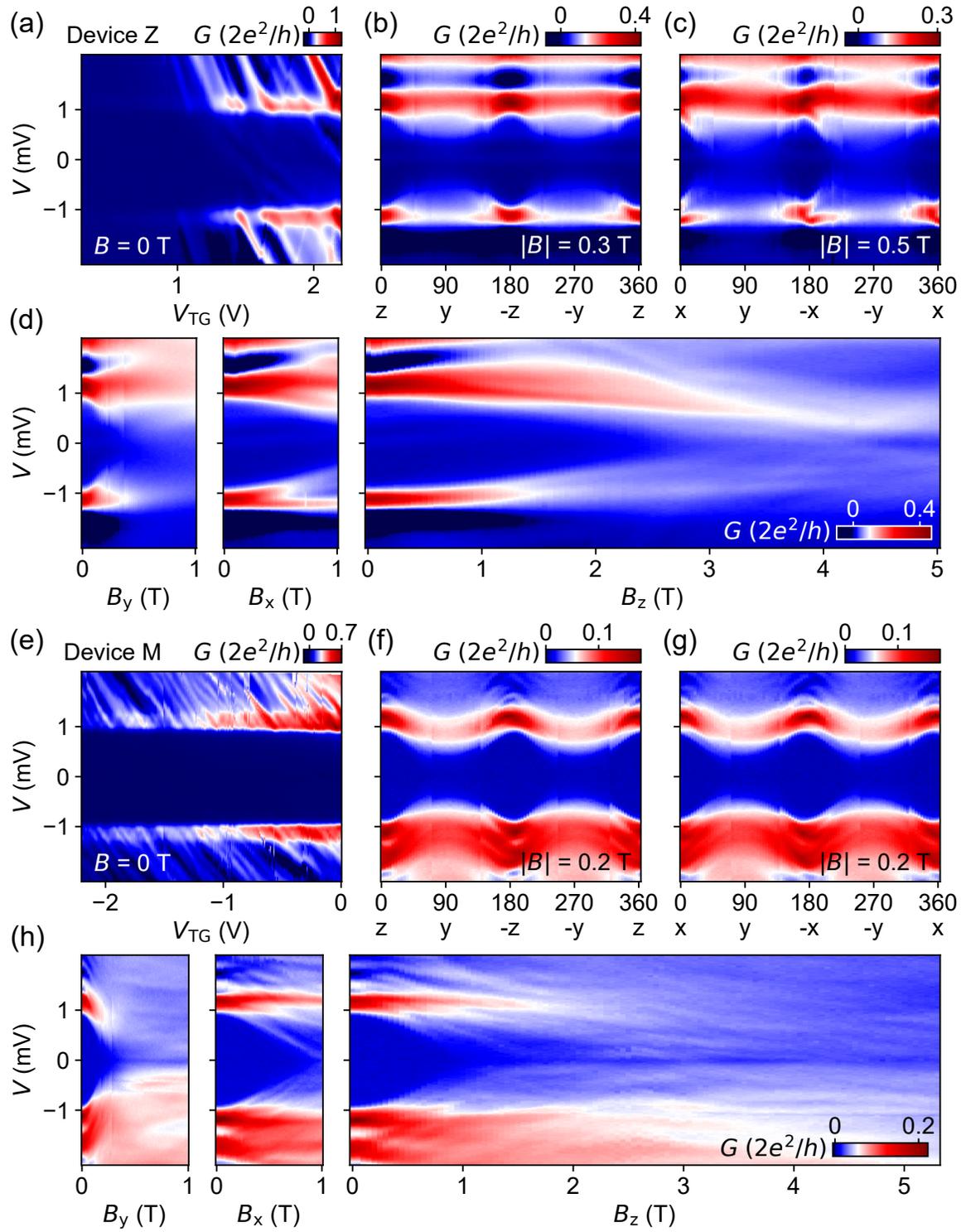

FIG. S6. Additional data of devices Z (a-d) and M (e-h). (a) Gap spectroscopy of device Z at zero field. $V_{SG} = 0$ V. (b-c) $B$ rotations in the $yz$ and $xy$ planes. $|B| = 0.3$ T and 0.5 T, respectively. $V_{TG} = 1.06$ V, $V_{SG} = 2$ V. (d) $B$ scans of the gap along $y$, $x$, and $z$ axes. (e-h) Similar to (a-d) but for device M.

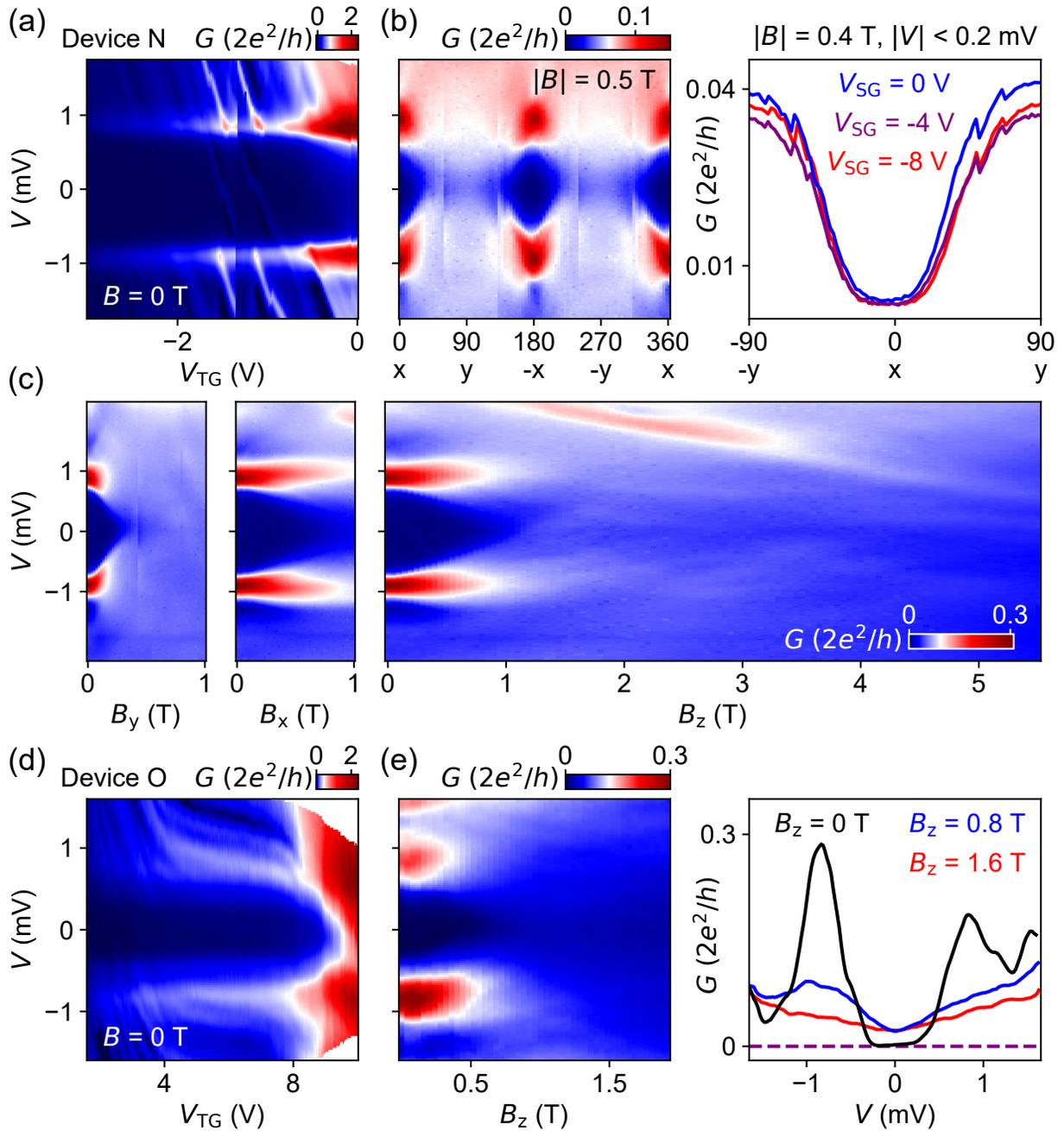

FIG. S7. Additional data of devices N (a-c) and O (d-e). (a) Gap spectroscopy of device N at zero field. $V_{SG} = 0$ V. (b) $B$ rotation in the $xy$ plane. $V_{TG} = -1.96$ V, $V_{SG} = 0$ V. $|B| = 0.5$ T. Right panel, averaged conductance near zero bias ($|V| < 0.2$ mV) for $V_{SG} = 0$ V (blue), -4 V (violet), and -8 V (red). The direction of minimal conductance barely changes for varying $V_{SG}$. (c) $B$ scans of the gap along $y$, $x$, and $z$ axes. (d) Gap spectroscopy of device O at zero field. $V_{SG} = 4$ V. (e) $B_z$ scan of the gap. $V_{TG} = -10.2$ V, $V_{SG} = 8$ V. Right panel, line cuts at 0, 0.8, and 1.6 T.


\end{document}